\documentclass[12pt,oneside]{article}
\usepackage{graphicx}
\usepackage{epstopdf}
\def\u1p{U(1)$^{\prime}$}
\def\zp{Z$^{\prime}$ }
\newcommand{\be}{\begin{equation}}
\newcommand{\ee}{\end{equation}}
\newcommand{\bea}{\begin{eqnarray}}
\newcommand{\eea}{\end{eqnarray}}

\newcommand{\bth}{{\bf 3}}
\newcommand{\btw}{{\bf 2}}
\newcommand{\bon}{{\bf 1}}
\def\simlt{\mathrel{\raise.3ex\hbox{$<$\kern-.75em\lower1ex\hbox{$\sim$}}}}
\def\simgt{\mathrel{\raise.3ex\hbox{$>$\kern-.75em\lower1ex\hbox{$\sim$}}}}
\begin{document}

%------------------------TITLE-----------------------------
\begin{titlepage}
\begin{flushright}
IZTECH-P02/2007
\end{flushright}
\vspace{0.5cm}
\begin{center}
{\Large \bf Dilepton Signatures of Family Non-universal \u1p} \\
\vspace{1cm} {\large Alper Hayreter
\\ {\it
Department of Physics, Izmir Institute of Technology, IZTECH, \\
Izmir,  TR 35430, Turkey}}
\end{center}
\vspace{0.5cm}

%%%----------------------------------------------------------------------

\begin{abstract}
The supersymmetric models extending the minimal supersymmetric
standard model (MSSM) by an additional Abelian gauge factor \u1p
in order to solve the $\mu$ problem do generically suffer from
anomalies disrupting the gauge coupling unification found in the
MSSM. The anomalies are absent if the minimal matter content
necessitated by the $\mu$ problem is augmented with exotic matter
species having appropriate quantum numbers. Recently, it has been
shown that anomaly cancellation can also be accomplished by
introducing family non-universal \u1p charges and non-holomoprhic
soft-breaking terms [Ref. 11]. We discuss collider signatures of
anomaly-free family non-universal \u1p model by analyzing dilepton
production in future colliders. We find that, both at LHC and NLC,
one can establish existence/absence of such a Z$^{\prime}$ boson
by simply comparing the number of dilepton production events for
electron, muon and tau lepton. The signal is free of the SM
background.

\end{abstract}
\vspace*{1cm}
\end{titlepage}

\section{Introduction}
The Minimal Supersymmetric Standard Model (MSSM), devised to solve
the gauge hierarchy problem of the standard model of electroweak
interactions (SM), suffers from a serious naturalness problem
associated with the Dirac mass of Higgsinos in the superpotential.
Namely, the dimensionful parameter $\mu$ contained in the
superpotential

\begin{eqnarray}
\label{wmssm} \widehat{W} \ni \mu \widehat{H}_u \cdot
\widehat{H}_d
\end{eqnarray}
%%%%%%%%%%%%%%%%%%%%%%%%%%%%%%%%%%%%%%%%%
is nested in the supersymmetric sector of the theory, and its
scale is left completely arbitrary as it is not related to the
soft supersymmetry-breaking terms \cite{muprob}. A way out of this
problem is to generate $\mu$ parameter dynamically via the vacuum
expectation value (VEV) of some SM-singlet chiral superfield. The
extension by a non-SM chiral superfield may or may not involve
gauge extension. Concerning the former, the most conservative
approach is to extend the gauge structure of the MSSM by an extra
Abelian group factor \u1p along with an additional chiral
superfield $\widehat{S}$ whose scalar component generates an
effective $\mu$ parameter upon spontaneous \u1p breakdown. What
this additional gauge symmetry actually does is to forbid the
presence of a bare $\mu$ parameter as in (\ref{wmssm}) \cite{gut,
string,dynamic}. An important property of \u1p models is that the
lightest Higgs boson weighs significantly heavier than $M_Z$ even
at tree level with small $\tan\beta$. Hence the existing LEP
bounds \cite{lep:2003,lep:2005} are satisfied with almost no need for large radiative
corrections \cite{Cvetic:1997ky,everett,han,amini}. Besides, they
offer a rather wide parameter space for facilitating the
electroweak baryogenesis \cite{tianjun}.

An important problem in \u1p models concerns the cancellation of
anomalies. Indeed, for making the theory anomaly--free the usual
approach to \u1p models is to add several exotics to the spectrum
\cite{Erler:2000wu}. This naturally happens in \u1p models
following from SUSY GUTs $e.g.$ E$_6$ unification. However, this
not only causes a significant departure from the minimal structure
but also disrupts the gauge coupling unification -- one of the
fundamental predictions of the MSSM with weak scale soft masses.
Therefore, it would be of greatest interest to keep gauge
unification with minimal matter content. This has been
accomplished in \cite{05makalesi} by introducing family
non-universal \u1p charges in a way solving all anomaly
conditions, including the gravitational one.

In this work, we will discuss dilepton signatures of \u1p models
with universal as well as non-universal \u1p charges in a
comparative fashion. Our discussion will include both lepton (the
ILC) and hadron (the LHC) colliders. At the Born level the cross
sections are sensitive to \zp exchange only. Therefore, our
analysis will have examined \zp properties via dilepton signal.
The collider signatures of various \u1p models have already been
analyzed in the literature \cite{collid,langacker-kang,spin}. In
addition, the \u1p models have also been tested under electroweak
precision bounds \cite{indirect}.

\section{The \u1p Model}
In \u1p models the MSSM gauge group is extended to include an
extra Abelian group factor at the weak scale:
SU(3)$_C\times$SU(2)$_L \times$U(1)$_Y\times$\u1p with respective
gauge couplings $g_3$, $g_2$, $g_1$ and $g_1^{\prime}$. The
particle spectrum of the model is that of the MSSM plus a MSSM
gauge singlet $S$ charged under only the \u1p invariance. We
employ a rather general \u1p charge assignment as tabulated in
Table \ref{table1}.

\begin{table}[h]
\begin{center}
\begin{tabular}{|c|c|c|c|c|} \hline
      & SU(3)$_c$ & SU(2)$_L$  & U(1)$_Y$ & \u1p   \\ \hline
$Q_i$   & $\bth$       &  $\btw$       &  $1/6$   & $Q^{\prime}_{Q_i}$ \\
\hline $U^c_i$   & $\bar \bth$  &  $\bon$       &  $-2/3$  &
$Q^{\prime}_{U^c_i}$ \\ \hline $D^c_i$   & $\bar \bth$  &  $\bon$
& $1/3$   & $Q^{\prime}_{D^c_i}$ \\ \hline $L_i$   & $\bon$ &
$\btw$ &  $-1/2$  & $Q^{\prime}_{L_i}$ \\ \hline $E^c_i$   &
$\bon$ & $\bon$ &  $1$     & $Q^{\prime}_{E^c_i}$ \\ \hline $H_u$
& $\bon$ & $\btw$ &  $1/2$   & $Q^{\prime}_{H_u}$ \\ \hline $H_d$
& $\bon$ &  $\btw$ & $-1/2$  & $Q^{\prime}_{H_d}$ \\ \hline $S$ &
$\bon$ &  $\bon$ & $0 $   & $Q^{\prime}_{S}$ \\ \hline
\end{tabular}
\end{center}
\caption{ The gauge quantum numbers of chiral superfields of
$i$-th family.} \label{table1}
\end{table}
%%%%%%%%%%%%%%%%%%%%%%%%%%%%%%%%%%%%%%%%%%%%%%%%%%%%%
As shown in \cite{05makalesi}, this general \u1p charge assignment
suffices to solve all anomaly cancellation conditions in a way
respecting the gauge invariance of the superpotential. In fact,
one finds the solutions \cite{05makalesi}

\begin{eqnarray}
Q^{\prime}_{Q_1}&=&Q^{\prime}_{Q_2}=Q^{\prime}_{Q_3}=\frac{1}{9}(3Q^{\prime}_{E^c_2}+3Q^{\prime}_{L_2}+Q^{\prime}_S)\,,
\nonumber\\
Q^{\prime}_{D^c_1}&=&Q^{\prime}_{D^c_2}=Q^{\prime}_{D^c_3}=\frac{1}{9}(6Q^{\prime}_{E^c_2}+6Q^{\prime}_{L_2}-Q^{\prime}_S)\,,
\nonumber\\
Q^{\prime}_{U^c_1}&=&Q^{\prime}_{U^c_2}=Q^{\prime}_{U^c_3}=\frac{1}{9}(-12Q^{\prime}_{E^c_2}-12Q^{\prime}_{L_2}-Q^{\prime}_S)\,,
\nonumber\\ Q^{\prime}_{L_1}&=&-2Q^{\prime}_{E^c_2}-3Q^{\prime}_{L_2}\,,\;\,Q^{\prime}_{L_3}=-Q^{\prime}_{E^c_2}-Q^{\prime}_{L_2}\,, \nonumber\\
Q^{\prime}_{E^c_1}&=&3Q^{\prime}_{E^c_2}+4Q^{\prime}_{L_2}\,,\;\,Q^{\prime}_{E^c_3}=2Q^{\prime}_{E^c_2}+2Q^{\prime}_{L_2}+Q^{\prime}_S\,,\nonumber\\
Q^{\prime}_{H_d}&=&-Q^{\prime}_{E^c_2}-Q^{\prime}_{L_2}-Q^{\prime}_S\,,\;\,Q^{\prime}_{H_u}=Q^{\prime}_{E^c_2}+Q^{\prime}_{L_2}
\label{noanomal}
\end{eqnarray}
%%%%%%%%%%%%%%%%%%%%%%%%%%%%%%%%%%%%%%%%%%%%%%%%%%%%
in terms of the three free charges:

\begin{eqnarray}
\label{choice}
Q^{\prime}_{L_2}=2\;,\;\;Q^{\prime}_{E^c_2}=-3\;,\;\;
Q^{\prime}_S=3\,,
\end{eqnarray}
%%%%%%%%%%%%%%%%%%%%%%%%%%%%%%%%%%%%%%%%%%%%%%%%%%%
in the theory, these free charges are normalized with a factor
$C_{Z^{\prime}}$ which varies with the normalizing model.

In what follows, we will use this general solution of the charges
in analyzing the collider signatures of family non-universal \u1p.
The theory consists of three gauge bosons: the photon, the Z boson
and the \zp boson. We parameterize couplings of these vector
bosons to fermions via the effective lagrangian \cite{quiros}:

\begin{eqnarray}
\label{lagran} {\cal{L}}_{eff} = \frac{g_2}{4 \cos \theta_W}
\sum_{i} \overline{f}_i \gamma^{\mu} \left( v_{V}^{f} - a_{V}^{f}
\gamma^5\right) f_i \mbox{V}_{\mu}
\end{eqnarray}
%%%%%%%%%%%%%%%%%%%%%%%%%%%%%%%%%%%%%%%%%%%%%%%%%%%%%%%
where $V=\gamma, \mbox{Z}, \mbox\zp$, and $f_i$ stands for any of
the quarks or leptons. The \u1p gauge coupling $g_1^\prime$ is
included in the vector couplings  $v_{V}^{f}$ and axial-vector
couplings $a_{V}^{f}$ via the relations
\begin{eqnarray}
\label{vector-axial-1}v_{V}^{f} = 2 \cos \theta_W \left(
Q^{\prime}_{f_L}-Q^{\prime}_{f_R} \right) \frac{g_1^\prime}{g_2}
\;\;\;\;\;,\;\;\;\;\; a_{V}^{f} = 2 \cos \theta_W \left(
Q^{\prime}_{f_L}+Q^{\prime}_{f_R} \right)
\,\frac{g_{1}^\prime}{g_{2}}
\end{eqnarray}
%%%%%%%%%%%%%%%%%%%%%%%%%%%%%%%%%%%%%%%%%%%%%%%%%%%%%%%
where $\theta_W$ is the Weinberg angle, and $Q^{\prime}_{f_L}$ and
$Q^{\prime}_{f_R}$ are \u1p charges of left-- and right--handed
fermions, respectively.

\begin{table}
\begin{center}
\begin{tabular}{|c|c|c|c|c|c|c|}
  \hline
           & \multicolumn{2}{c|}{$\gamma$} & \multicolumn{2}{c|}{Z} & \multicolumn{2}{c|}{\zp} \\ \hline
           & $v$ & $a$ & $v$ & $a$ & $v$ & $a$ \\ \hline
  $\nu_e,\nu_{\mu},\nu_{\tau}$ & $0$ & $0$ & $1$ & $1$ & $-\sin{\theta}_W/3$ & $-\sin{\theta}_W/3$ \\ \hline
  $e^-,\mu^-,\tau^-$ & $-1$ & $0$ & $-1\,+\,4\,\sin^2{\theta}_W$ & $-1$ & $-\sin{\theta}_W$ & $\sin{\theta}_W/3$ \\ \hline
  $u,\,c,\,t$     & $2/3$ & $0$ & $1\,-\,8\,\sin^2{\theta}_W/3$ & $1$ & $0$ & $4\sin{\theta}_W$ \\ \hline
  $d,\,s,\,b$     & $-1/3$ & $0$ & $-1\,+\,4\,\sin^2{\theta}_W/3$ & $-1$ & $\sin{\theta}_W$ & $\sin{\theta}_W/3$ \\ \hline
\end{tabular}
\end{center}
\caption{The vector boson couplings to fermions with family
universal \u1p. The \u1p couplings here are those of U(1)$_{\eta}$
descending from E(6) supersymmetric GUT (see
\cite{langacker-kang}).} \label{table2}
\end{table}

\begin{table}
\begin{center}
\begin{tabular}{|c|c|c|c|c|c|c|}
  \hline
           & \multicolumn{2}{c|}{$\gamma$} & \multicolumn{2}{c|}{Z} & \multicolumn{2}{c|}{\zp} \\ \hline
           & $v$ & $a$ & $v$ & $a$ & $v$ & $a$ \\ \hline
  $\nu_e$ & $0$ & $0$ & $1$ & $1$ & $2\sin{\theta}_W\,C_{Z^{\prime}}$ & $-2\sin{\theta}_W\,C_{Z^{\prime}}$ \\ \hline
  $\nu_{\mu}$ & $0$ & $0$ & $1$ & $1$ & $10\sin{\theta}_W\,C_{Z^{\prime}}$ & $-2\sin{\theta}_W\,C_{Z^{\prime}}$ \\ \hline
  $\nu_{\tau}$ & $0$ & $0$ & $1$ & $1$ & $0$ & $4\sin{\theta}_W\,C_{Z^{\prime}}$ \\ \hline
  $e^-$ & $-1$ & $0$ & $-1\,+\,4\,\sin^2{\theta}_W$ & $-1$ & $2\sin{\theta}_W\,C_{Z^{\prime}}$ & $-2\sin{\theta}_W\,C_{Z^{\prime}}$ \\ \hline
  $\mu^-$ & $-1$ & $0$ & $-1\,+\,4\,\sin^2{\theta}_W$ & $-1$ & $10\sin{\theta}_W\,C_{Z^{\prime}}$ & $-2\sin{\theta}_W\,C_{Z^{\prime}}$ \\ \hline
  $\tau^-$ & $-1$ & $0$ & $-1\,+\,4\,\sin^2{\theta}_W$ & $-1$ & $0$ & $4\sin{\theta}_W\,C_{Z^{\prime}}$ \\ \hline
  $u,c,t$     & $2/3$ & $0$ & $1\,-\,8\,\sin^2{\theta}_W/3$ & $1$ & $-2\sin{\theta}_W\,C_{Z^{\prime}}$ & $2\sin{\theta}_W\,C_{Z^{\prime}}$ \\ \hline
  $d,s,b$     & $-1/3$ & $0$ & $-1\,+\,4\,\sin^2{\theta}_W/3$ & $-1$ & $2\sin{\theta}_W\,C_{Z^{\prime}}$ & $-2\sin{\theta}_W\,C_{Z^{\prime}}$ \\ \hline
\end{tabular}
\end{center}
\caption{The vector boson couplings to fermions with family
non-universal \u1p. The \u1p charges are determined by using
(\ref{choice}) and by the normalization condition that
$g_1^{\prime\, 2} \mbox{Tr}[Q^{\prime\, 2}]$ to be equal to the
same quantity computed in U(1)$_{\eta}$ model and the
normalization factor $C_{Z^{\prime}}$ is evaluated as
$\sqrt{\frac{5}{52}}$.} \label{table3}
\end{table}
%%%%%%%%%%%%%%%%%%%%%%%%%%%%%%%%%%%%%%%%%%%%%%%%%%%%%%
In writing (\ref{lagran}) we have neglected the mixing between Z
and \zp bosons. This mixing can stem from kinetic mixing or can be
induced after electroweak breaking \cite{kolda,Cvetic:1997ky}.In
this work we neglect such mixings in accord with the experimental
bounds that $\alpha_{\mbox{Z}-\mbox{Z}^{\prime}}$ cannot exceed a
few $10^{-3}$. This smallness of the mixing puts stringent bounds
on the ranges of the soft-breaking masses as it was analyzed in
detail in \cite{Cvetic:1997ky,05makalesi}.

\section{Dilepton Signatures of \u1p}
In this section we will analyze the family non-universal \u1p
model by considering its signatures for dilepton production at
lepton and hadron colliders, separately. We will investigate
distinctive signatures of the \u1p model under concern with
respect to a typical family universal \u1p model which we choose
to be the U(1)$_{\eta}$ model following from E(6) GUT. The
requisite vector and axial-vector couplings of photon, Z and \zp
bosons are tabulated in Table \ref{table2} and \ref{table3} for
family universal and non-universal models, respectively.

In general, the $2 \rightarrow 2$ scattering process
\begin{eqnarray}
\label{process} f\, \overline{f} \rightarrow \ell^+ \ell^-
\end{eqnarray}
where $f$ stands for quarks (hadron colliders) or leptons (lepton
colliders) and $\ell$ for any of the charged leptons. This process
proceeds with $\gamma$, Z and \zp exchanges in the $s$-channel
when $\ell$ is not identical to $f$, and in both $s$ and $t$
channels when $f\equiv \ell$. If center of mass energy of the
collider is high enough then \zp effects can be disentangled from
those of $\gamma$ and Z.

After summing over final-state polarizations and averaging over
the initial-state ones, the amplitude-squared of (\ref{process})
takes the form
\begin{eqnarray}
\langle\,\,|{\cal{A}}\left(f\, \overline{f} \rightarrow \ell^+
\ell^-\right)|^2\rangle_{polar.} \,=\, F(s; v,a)\,[(s+t)^2+t^2]\,
+ \,G(s; v,a)\,[(s+t)^2-t^2]
\end{eqnarray}
where $F(s; v,a)$ and $G(s; v,a)$ are given by \cite{quiros}
\begin{eqnarray}
F(s;
v,a)=2\sum_{\alpha,\beta}\frac{(v_{\alpha}^{f}\,v_{\beta}^{f}+a_{\alpha}^{f}\,a_{\beta}^{f})\,
(v_{\alpha}^{l}\,v_{\beta}^{l}+a_{\alpha}^{l}\,a_{\beta}^{l})}{(s-M_{\alpha}^2+iM_{\alpha}\Gamma_{\alpha})
(s-M_{\beta}^2-iM_{\beta}\Gamma_{\beta})}\nonumber
\end{eqnarray}
and
\begin{eqnarray}
G(s;
v,a)=2\sum_{\alpha,\beta}\frac{(v_{\alpha}^{f}\,a_{\beta}^{f}+
v_{\beta}^{f}\,a_{\alpha}^{f})\,(v_{\alpha}^{l}\,a_{\beta}^{l}+v_{\beta}^{l}\,a_{\alpha}^{l})}
{(s-M_{\alpha}^2+iM_{\alpha}\Gamma_{\alpha})(s-M_{\beta}^2-iM_{\beta}\Gamma_{\beta})}\,.
\end{eqnarray}
In these expressions $\alpha$ and $\beta$ label intermediate
vector bosons $i.e.$ ${\gamma}$, Z and  \zp. The $\Gamma_{\alpha}$
designates widths of the vector bosons: $\Gamma_{\gamma} = 0$
(absolutely stable) and $\Gamma_{\mbox{Z}} = 2.4952\ {\rm GeV}$.
The \zp width $\Gamma_{\mbox{Z}^{\prime}}$ is a model-dependent
quantity, and while making numerical estimates in what follows we
will take $\Gamma_{\mbox{Z}^{\prime}}= \Gamma_{\mbox{Z}}$.
Moreover, in accord with the U(1)$_{\eta}$ model parameter space,
we take $g_1^{\prime} = g_1$.

\subsection{The Linear Collider Signatures}
We first examine \u1p model at a high-energy linear collider (such
as the International Linear Collider (ILC) project under
preparation) running at $\sqrt{s}=500\ {\rm GeV}$. The basic
processes we consider are $e^+ e^- \rightarrow \mu^+ \mu^-$ and
$e^+ e^- \rightarrow \tau^+ \tau^-$ where we discard $e^+ e^-$
final states simply for avoiding the $t$-channel contributions.

\begin{figure} [htbp]
\begin{minipage}{8in}
\includegraphics[height=2in]{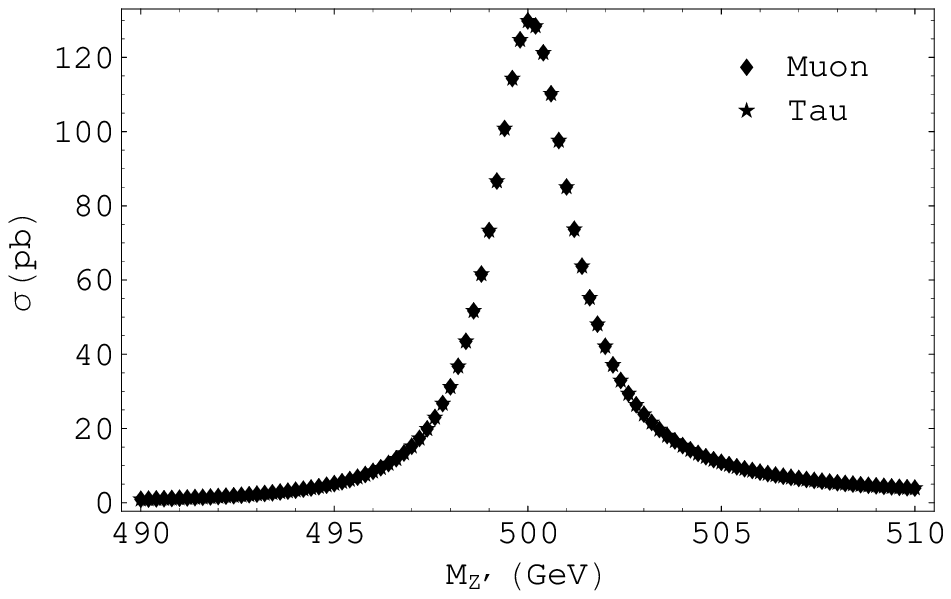}
\includegraphics[height=2in]{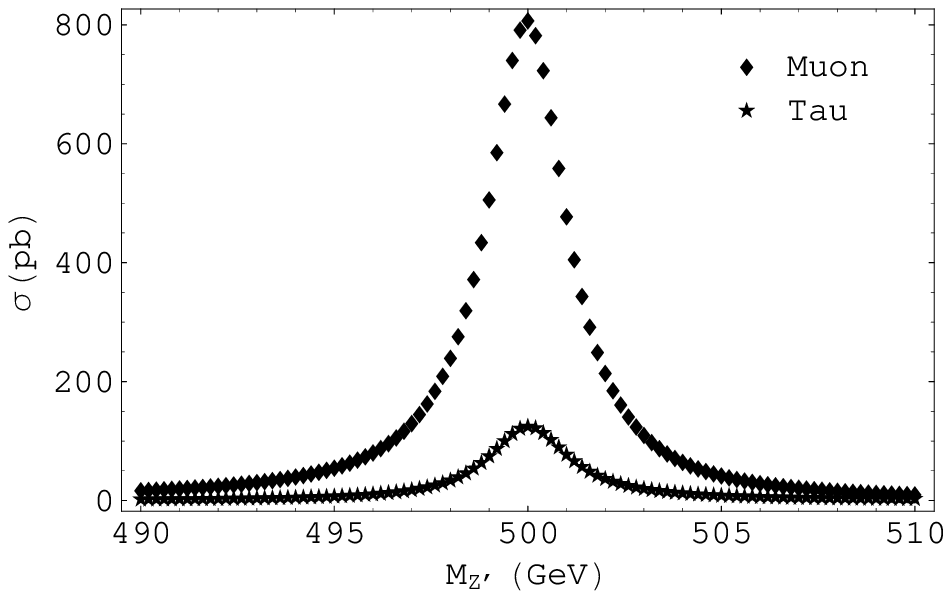}
\end{minipage}
\caption{The $\mu^{+}\mu^{-}$ and $\tau^{+}\tau^{-}$ productions
at a future $e^+ e^-$ collider with $\sqrt{s} = 500\ {\rm GeV}$
for family universal \u1p (in the left panel) and family
non-universal \u1p (in the right panel) models. The ratio between
family non-universal and family universal cross sections varies with
model parameters.} \label{fig1}
\end{figure}

Depicted in Fig. \ref{fig1} are unpolarized $\mu^{+}\mu^{-}$ and
$\tau^{+}\tau^{-}$ production cross sections
\begin{eqnarray}
\sigma\left(e^+ e^- \rightarrow \ell^+ \ell^-\right)=\frac{1}{16
\pi s} \int_{-s}^0 dt\, \langle\,\,|{\cal{A}}\left(f\,
\overline{f} \rightarrow \ell^+ \ell^-\right)|^2\rangle_{polar.}
\end{eqnarray}
at a future $e^+ e^-$ machine for family universal \u1p (in the
left panel) and family non-universal \u1p (in the right panel)
models. For family universal \u1p it is seen that $\sigma(e^+ e^-
\rightarrow \mu^{+}\mu^{-})$ and $\sigma(e^+ e^- \rightarrow
\tau^{+}\tau^{-})$ completely overlap. The main reason behind this
coincidence is that $\mu$ and $\tau$ leptons do have identical
gauge quantum numbers (including those of under the \u1p gauge
symmetry) and their mass difference causes only a tiny deviation
at such high energies \cite{lep:2003,lep:2005}. Consequently, from the left
panel of Fig. \ref{fig1} one concludes that numbers of muons and
tau leptons produced at an $e^+ e^-$ collider will be identical
(up to systematic and statistical errors in analyzing the
experimental data) if the new gauge symmetry, the \u1p symmetry
under concern, exhibits identical \zp couplings for each fermion
(at least lepton) family as happens in the standard electroweak
theory.

In clear contrast to the left-panel of Fig. \ref{fig1}, one
observes that $\mu^{+}\mu^{-}$ and $\tau^{+}\tau^{-}$ differ by an
order of magnitude if the \u1p symmetry possesses non-universal
couplings to fermions (at least leptons). Indeed, $\sigma(e^+ e^-
\rightarrow \mu^{+}\mu^{-})$ is larger than $\sigma(e^+ e^-
\rightarrow \tau^{+}\tau^{-})$ by a factor of $6.5$, and this
factor is related to \u1p charges listed in Table \ref{table1} and
vector and axial-vector couplings in Table \ref{table3}.
Therefore, the right-panel of Fig. \ref{fig1} alone is sufficient
for concluding that the number of $\mu^{+}\mu^{-}$ and
$\tau^{+}\tau^{-}$ events will significantly differ from each
other if the new gauge symmetry, the \u1p gauge symmetry under
concern, exhibits different \zp couplings to different fermion (at
least lepton) families.

\begin{figure} [htbp]
\begin{minipage}{6.5in}
\begin{center}
\includegraphics[height=2in]{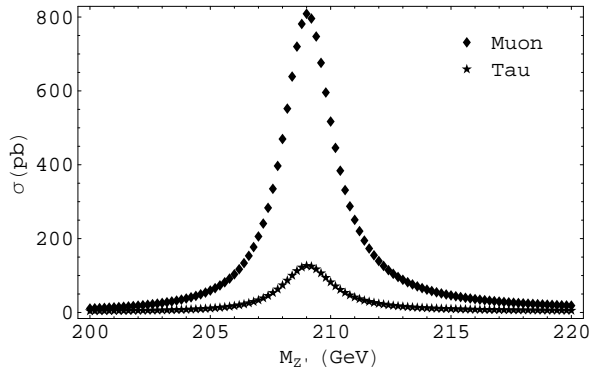}
\end{center}
\end{minipage}
\caption{ Family non-universal \zp at LEP }\label{fig2}
\end{figure}

%%%%%%%%%%%%%%%%%%%%%%%%%%%%%%%%%%%%%%%%%%%%%%%%%%%
Additionally we analyze the \u1p model at the Large Electron-Positron (LEP) collider which is closed at 2000 with $\sqrt{s}=209\ {\rm GeV}$ and 140$\ {\rm pb}^{-1}$ luminosity. Fig. \ref{fig2} is the production cross sections of  muon and tau lepton final states with family non-universal \u1p. It is clear in Fig.\ref{fig2} that family non-universal \u1p signal is quite clean and distinguishable as the muon and tau lepton production cross sections are as much as several hundreds of picobarns. However, these productions are observed to be around few picobarns in various analysis \cite{collid,lep:2003,lep:2005} and since such a clear and distinct signal has not been observed in LEP \cite{lep:2003,lep:2005}, it can easily be said that family non-universal \zp lies beyond the discovery limit of LEP.

In conclusion, at linear colliders, which provide a perfect arena
for precision measurements, one can determine if the new gauge
symmetry, if any, which extends the SM gauge group exhibits family
universal or non-universal couplings by simply counting the number
of lepton pairs produced. This aspect is quite important since
family non-universality might signal anomaly cancellation in
Abelian extended models as shown in \cite{05makalesi}.

\subsection{The Hadron Collider Signatures}

The most important hadron machine to come up is the Large Hadron
Collider (LHC) which is a proton-proton collider running at
$\sqrt{s} = 14\ {\rm TeV}$ center of mass energy. At the parton
level dilepton production processes are started by
quark--anti-quark annihilation into lepton pairs via $s$-channel
$\gamma$, Z and \zp exchanges.

The hadronic cross section is related to the partonic one via
\begin{eqnarray}
\sigma\left(pp\rightarrow \ell^+ \ell^-\right)
=\sum_{q,\bar{q}}C_{q\bar{q}}\int
dx_{q}\,dx_{\bar{q}}\,{\cal{P}}_{q/A}(x_{q}){\cal{P}}_{\bar{q}/B}(x_{\bar{q}})\,
\sigma\left(q\bar{q}\rightarrow \ell^+ \ell^- \right)
\end{eqnarray}
%%%%%%%%%%%%%%%%%%%%%%%%%%%%%%%%%%%%%%%%%%%%%%%%%%%
where ${\cal{P}}_{q/A}(x_q)$ stands for probability of finding
parton (quark) $q$ within the hadron $A$ with a longitudinal
momentum $x_q$ time that of the hadron. Moreover, $C_{q\bar{q}}$
stands for color averaging over initial-state partons and it
equals $1/9$ for $q\, \bar q$ annihilation.

\begin{figure} [htbp]
\begin{minipage}{8in}
\includegraphics[height=2in]{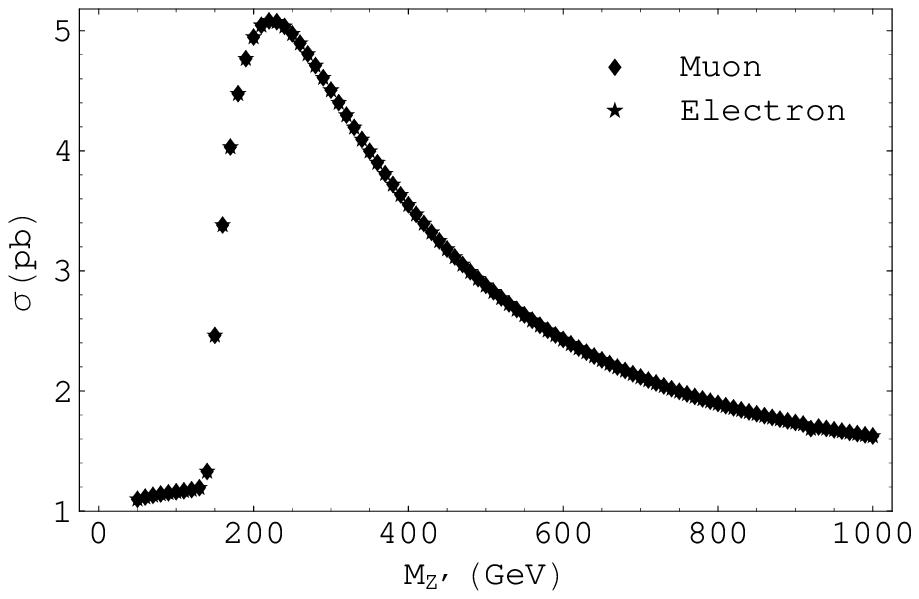}
\includegraphics[height=2in]{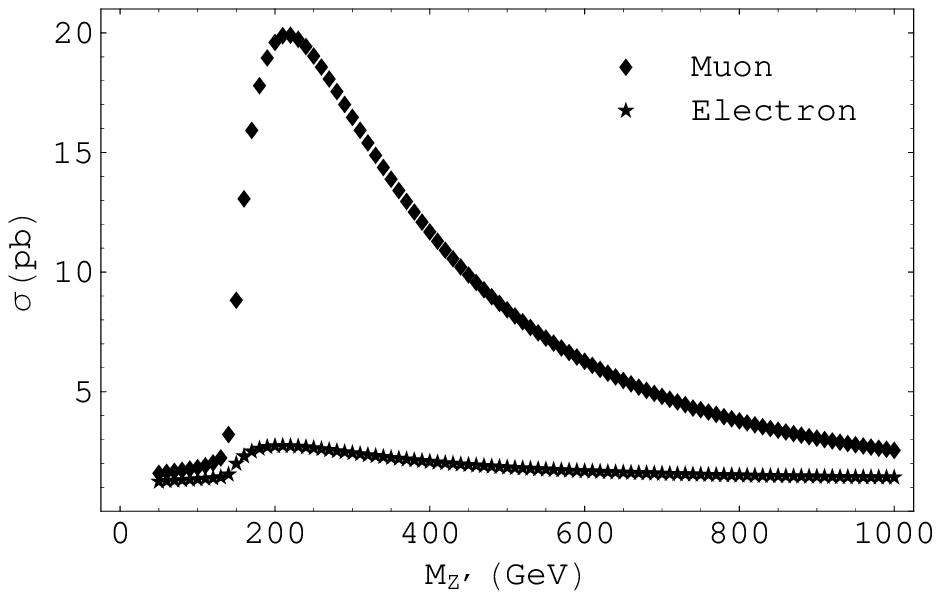}
\end{minipage}
\caption{ The unpolarized $e^+ e^-$ and $\mu^+ \mu^-$ productions
at the LHC for family universal (left panel) and non-universal (in
the right panel) \u1p models. The ratio between family non-universal
and family universal cross sections varies with model parameters.}\label{fig3}
\end{figure}

Depicted in Fig. \ref{fig3} are $\sigma\left(pp\rightarrow e^+
e^-\right)$ and $\sigma\left(pp\rightarrow \mu^+ \mu^-\right)$ for
family universal (in the left panel) and non-universal (in the
right panel) models. From the left-panel it is clear that the two
cross sections coincide, that is, an additional \u1p symmetry with
universal couplings to fermion (at least lepton) families is
expected to lead equal numbers of $e^+ e^-$ and $\mu^+ \mu^-$
pairs at the LHC. This observation is similar to what we found
while analyzing ILC signatures in Sec. 3.1 above because of the
fact that U(1)$_{\eta}$ model possesses family universal couplings
and mass difference between muon and electron cannot induce an
observable effect on cross sections at such a high-energy collider
\cite{lep:2003,lep:2005}.

Similar to the right-panel of Fig. \ref{fig1}, the right-panel of
Fig. \ref{fig3} shows $e^+ e^-$ and $\mu^+ \mu^-$ production cross
sections at the LHC with family non-universal \u1p model. The
panel manifestly shows that $\sigma\left(pp\rightarrow e^+
e^-\right)$ is approximately $13$ times smaller than
$\sigma\left(pp\rightarrow \mu^+ \mu^-\right)$ because of unequal
\u1p charges of electron and muon tabulated in Table \ref{table1}
as well as their vector and axial-vector couplings given in Table
\ref{table3}. Therefore, a family non-universal \u1p, if any, can
have observable signatures at the LHC via dilepton production
processes.

\begin{figure} [htbp]
\begin{minipage}{6.5in}
\begin{center}
\includegraphics[height=2in]{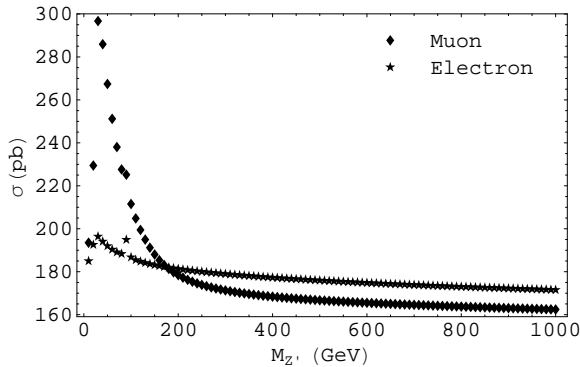}
\end{center}
\end{minipage}
\caption{ Family non-universal \zp at Tevatron }\label{fig4}
\end{figure}

%%%%%%%%%%%%%%%%%%%%%%%%%%%%%%%%%%%%%%%%%
We also examine the family non-universal \u1p model at $p-\bar{p}$ collisions with current bounds from Tevatron ($\sqrt{s}=2\ {\rm TeV}$). Fig. \ref{fig4} shows muon and electron production cross sections at Tevatron with family non-universal \u1p. Nevertheless the CDF \cite{Abe:1997,Abe:1995,Abe:1992} and D0 \cite{Abachi:1996,Abazov:2001,Abbott:1998} experiments are expected to probe \zp roughly in the range of 200-800 GeV masses for various models, thus Tevatron experiments put strong limits on \zp masses in agreement with the limits set by the LEP experiments. As it is understood in Fig.\ref{fig4}, family non-universal \u1p by being out of the limits is excluded at Tevatron with current bounds.
 
Before closing this section, we put strong emphasis on the fact
that family non-universal \u1p offers observable signatures in
dilepton signal in both linear and hadron colliders. In this
sense, the LHC, which is expected to start operation in coming
years, will be able to establish existence/absence of an
additional \u1p symmetry in general and a family non-universal
\u1p in particular. The latter will have easier observational
characteristics because all that matters is the measurement of the
ratios of events with different lepton flavors.

\section{Conclusion and Outlook}
In this work we have contrasted family universal and non-universal
\u1p models via their dilepton signatures in future linear (the
ILC) and hadron (the LHC) colliders. These production signatures are also 
observable in current colliders, and there are more stringent bounds on \zp 
from precision electroweak experiments and from direct searches in LEP \cite{lep:2003,lep:2005} and Tevatron \cite{Abe:1997,Abe:1995,Abe:1992,Abachi:1996,Abazov:2001,Abbott:1998}. 
The limits are model dependent because of the different couplings to fermions but typically 
the mass of a light \zp is comparable with Z ($\sim{\rm{200\, GeV}}$) and the heavy one is around 500-800 GeV with small mixings \cite{Cvetic:1997ky,Langacker:2003tv,lep:2003,lep:2005}.

Fig. \ref{fig2} and Fig. \ref{fig4} can be used in comparison between current and future colliders. Similar to ILC analysis Fig. \ref{fig2} indicates a family non-universal \u1p model with current bounds in LEP and Fig. \ref{fig4} is family non-universal \u1p at Tevatron in a similar fashion with LHC analysis. And again the family non-universality is at the difference in production cross sections of different flavors. As a result, family non-universal \zp is out of limits set by various experiments in LEP and Tevatron.

From discussions in Sec. 3.1 and Sec. 3.2 we conclude that in both colliders 
(depending on systematic and statistical error bars in experimental data) 
one can establish existence/absence of a family non-universal \u1p
model. This search is actually easier than direct \zp search since
all that matters is the ratio of production cross sections of
different lepton flavors.

For having a clearer sense of \zp search at colliders, it would be
useful to analyze decay patterns of \zp boson into different
flavors of matter. In general, a \zp boson of mass
$M_{\mbox{Z}^{\prime}}$ decays into a fermion $f$ and anti-fermion
$\overline{f}$ with a rate
\begin{eqnarray}
\Gamma_{Z^{\prime} \rightarrow f \bar{f}} = M_{\mbox{Z}^{\prime}}
\left(\frac{g_2}{4\cos\theta_W}\right)^2
\left(\frac{{v_{Z^{\prime}}^f}^2+{a_{Z^{\prime}}^f}^2}{12\pi}\right)
\end{eqnarray}
%%%%%%%%%%%%%%%%%%%%%%%%%%%%%%%%%%%%%%%%%%%%%%%%
directly proportional to $M_{\mbox{Z}^{\prime}}$. Therefore, if a
certain number of \zp bosons are produced (\zp bosons can be
copiously produced at the LHC) then their decays into different
fermion pairs gives information about the underlying structure of
the \u1p model.

Indeed, one expects at all grounds
\begin{eqnarray}
\frac{\Gamma_{Z^{\prime} \rightarrow \mu+\mu-}}{\Gamma_{Z^{\prime}
\rightarrow
\tau+\tau-}}\,=1\;\;\;\;\;;\;\;\;\;\;\frac{\Gamma_{Z^{\prime}
\rightarrow \mu+\mu-}}{\Gamma_{Z^{\prime} \rightarrow e+e-}}\,=1
\end{eqnarray}
in any \u1p model (may it follow from E(6) or from strings) in
which \zp couples to each lepton family in a universal fashion.

However, the same ratios of the decay rates become
\begin{eqnarray}
\frac{\Gamma_{Z^{\prime} \rightarrow \mu+\mu-}}{\Gamma_{Z^{\prime}
\rightarrow
\tau+\tau-}}\,=\,6.5\;\;\;\;\;;\;\;\;\;\;\frac{\Gamma_{Z^{\prime}
\rightarrow \mu+\mu-}}{\Gamma_{Z^{\prime} \rightarrow
e+e-}}\,=\,13
\end{eqnarray}
in the \u1p model of \cite{05makalesi} in which \zp couples to
different lepton families differently (as listed in Tables
\ref{table1} and \ref{table3}). That the decay rates can
significantly (depending on the model parameters) deviate from
unity is a highly interesting signature for collider searches for
a family non-universal \u1p gauge symmetry.

From the analyzes presented above we conclude that a \u1p gauge
symmetry with non-universal couplings to lepton families offers
unique observational signatures for collider searches via dilepton
production.

\section{Acknowledgements}

The author thanks D. A. Demir for suggesting this project and
several discussions and conversations while studying it. He is
grateful to Kerem Canko{\c c}ak for his numerical help and
guidance at various stages of this work. The author
also thanks to participants of Ankara Winter Workshop, January
2007, for discussions and criticisms about this work. He gratefully
acknowledges Turkish Academy of Sciences for financial support
through the GEBIP grant (through D. A. Demir).

\end{document}